\documentclass[12pt]{article}
\usepackage{amssymb,amsmath}
\usepackage{natbib}
\usepackage{verbatim,graphicx}
\usepackage{hyperref}
\hypersetup{colorlinks=true,
 urlcolor=black,
citecolor=black,
linkcolor=black }
\usepackage[top=1in, left=1in, right =1in]{geometry}
\title{\Large{\textbf{Testing the Capital Asset Pricing Model (CAPM)\\ 
 on the Uganda Stock Exchange}}}
\author{\textbf{David Wakyiku} \\
African Institute for Mathematical Sciences, South Africa.\\
 E-mail: davidw@aims.ac.za\\}
\date{}

\begin{document}
\maketitle
\abstract{\noindent This paper examines the validity of the Capital Asset Pricing Model (CAPM) on the Ugandan stock market using monthly stock returns from 10 of the 11 companies listed on the Uganda Stock Exchange (USE), for the period 1st March 2007 to 10th November 2009. Due to the absence of readily available Uganda Stock Exchange (USE) data, and the placement of daily price lists in \texttt{pdf} only, on the USE website: \href{http://www.use.or.ug}{\texttt{http://www.use.or.ug}}, the article also discusses the procedures taken to mine the data needed. The securities were all put in one porfolio in order to diversify away the firm-specific part of returns thereby enhancing the precision of the beta estimates. This paper should be of interest to both Ugandan and non-Ugandan investors and market researchers.  While many developing countries have legal restrictions against foreign participation in capital and money markets, this is not so in Uganda, where it has become part of government policy to encourage foreign capital inflow, inorder to stimulate the development of the small and underdeveloped markets.
\\
\\
The \cite{bjs} CAPM version is examined in this article. This version predicts a non zero-beta rate, along with the relation of higher returns to higher risk. The estimated zero-beta rate obtained is not statistically different from zero, and the estimated portfolio beta coefficient is statistically significant, providing evidence that the traditional form of CAPM holds on the USE, albeit having a beta coefficient that is not good at explaining the relationship between risk and return.
}

\section{Introduction}
One of the most important developments in modern capital theory is the capital asset pricing model (CAPM) developed by \cite{sharpe64} and \cite{lintner65}. This model was the first apparently successful attempt to show how to assess the risk of the cash flow from a potential investment project and to estimate the project's cost of capital, the expected rate of return that investors will demand if they are to invest in the project. The CAPM was developed, at least in part, to explain the differences in risk premium across assets. According to the CAPM, these differences are due to differences in the riskiness of the returns on the assets. The model asserts that the correct measure of riskiness is known as \textit{beta}, and that the risk premium per unit of riskiness is the same across all assets. Given the risk-free rate and the beta of an asset, the CAPM predicts the expected risk premium for that asset. Although the CAPM has been predominant in empirical work over the past 30 years and is the basis of modern portfolio theory, accumulating research has increasingly cast doubt on its ability to explain the actual movements of asset returns. \cite{banz81} and \cite{famafrench} have raised important insufficies, and even though there have been counter arguments, a CAPM debate has ensued.\footnote{\cite{jaganna} and \cite{grigoris06} give a thorough discussion of the CAPM debate.}
\\
\\
The purpose of this paper is to examine the validity of the CAPM on the USE.  Tests were conducted for a 33-month period using monthly stock returns. The Uganda Stock Exchange (USE), which has been in existence for ten years now\footnote{The capital markets became active in 1999, with the IPO of Uganda Clays (UCL) stock that was 100\% government owned.} has had its All Share Index grow steadily from the 200s in 2003 to the 800s at the beginning of 2007, reaching its peak of 1162.49 on 10th June 2008, before slumping to the the 700s as a result of the ripple effects of the credit crunch. The period of March 2007 to November 2009, for which the tests were conducted, is characterized by a high level of returns volatility.
\\
\\
The focus of research on the USE has previously been the legal framework, stock markets contribution to economic development, and liquidity issues. \cite{lutwama06}, \cite{atuhairwe05}, \cite{katto04} and a few others, carry out non-econometric discussions of these issues. It is only \cite{mayanja07} who attempt an econometric analysis on the USE, and therein is a computation of stock betas using what \cite{mayanja07} call the \lq\lq covariance method\rq\rq,~where they compute beta using the sample covariance and variance for the period February 2006 to March 2007. However, these results are sample-period biased since they do not include the 2008 to 2009 period during which the USE ALSI has been trending downwards. Moreover, \cite{mayanja07} do not discuss all the issues involved in the regression procedure, ubiquitous in CAPM analysis. CAPM is also tested on individual stocks only, which makes the betas imprecise since the firm-specific part of risk is not diversified away, leaving their analysis devoid of the portfolio procedure advocated by \cite{blume}, \cite{bjs}~and \cite{famamacbeth}.
\\
\\
In this article, tests of the CAPM model were first carried out on individual stocks, and then on the portfolio consisting of all the stocks listed on the USE, using monthly returns. 
\section{Sample Selection and Data}
\subsection{Sample Selection}
\begin{figure}[h]
\begin{center}
\includegraphics{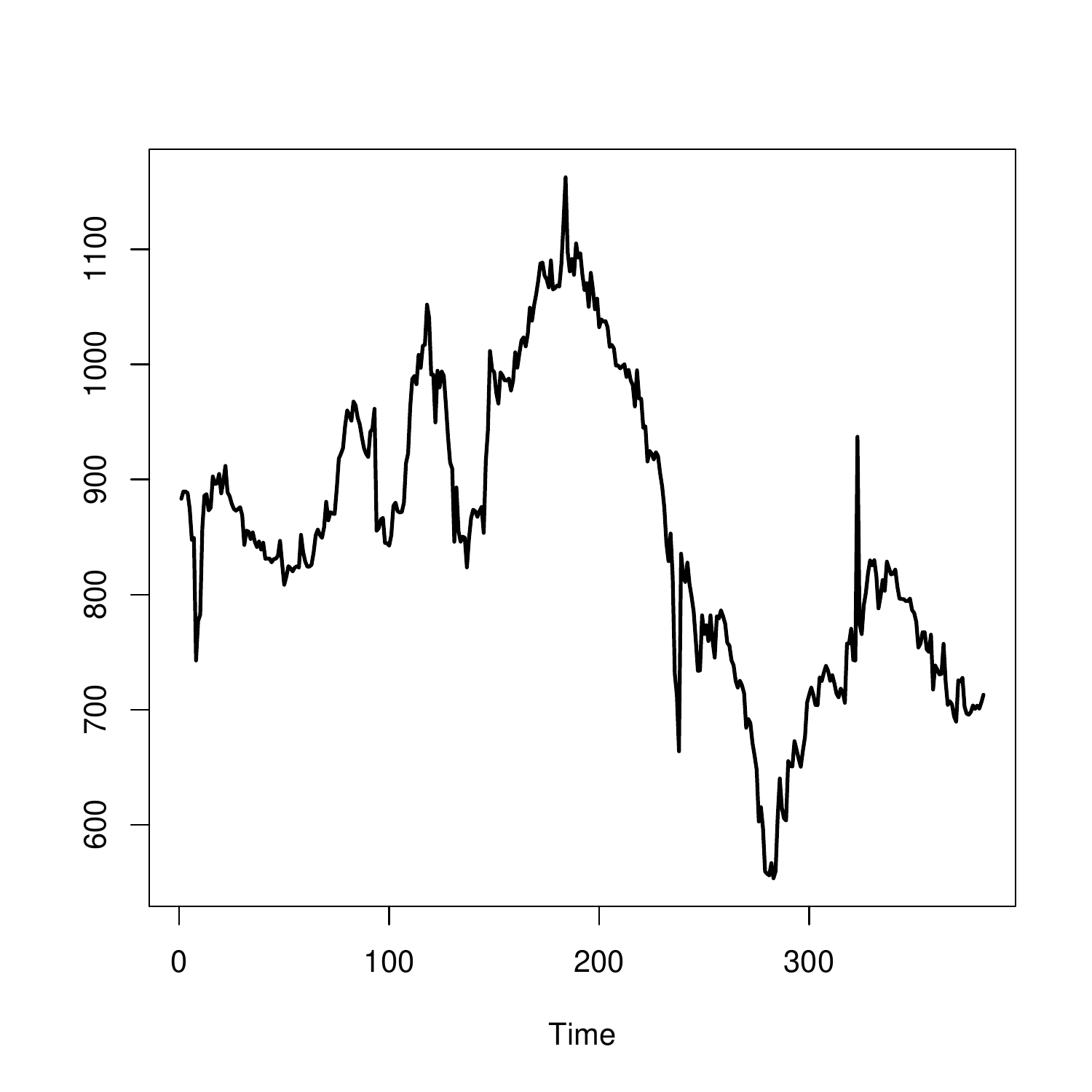}
\end{center}
\caption{USE ALSI for the period: 1st March, 2007 to 10th November, 2009.}
\label{fig:fig1}
\end{figure}

\begin{figure}[h!]
\begin{center}
\includegraphics{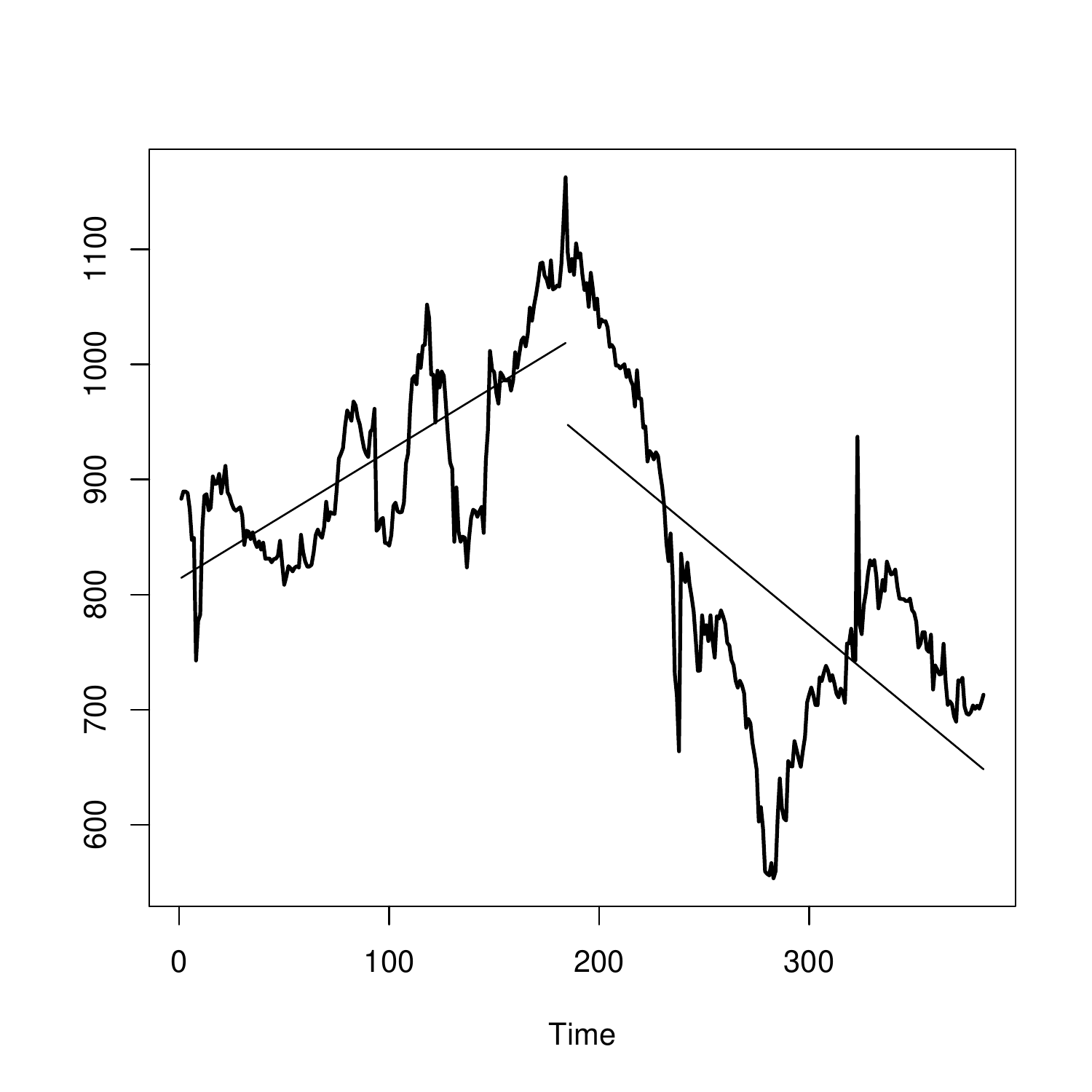}
\end{center}
\caption{USE ALSI for the period: 1st March, 2007 to 10th November, 2009, with inserted regression lines for the period before and after 10th June, 2008.}
\label{fig:fig2}
\end{figure}

\begin{figure}[h!]
\begin{center}
\includegraphics{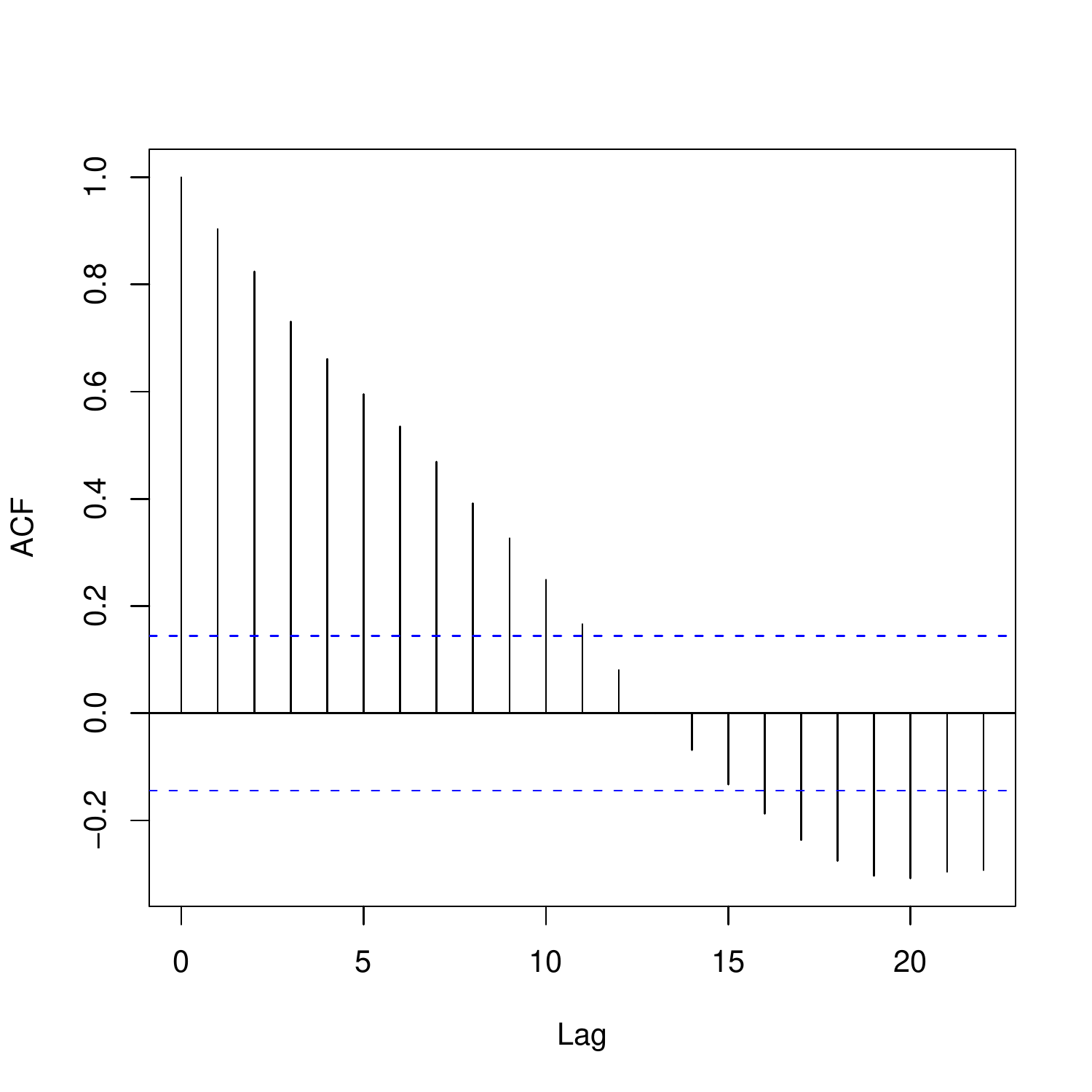}
\end{center}
\caption{Autocorrelation of regression residuals for ALSI on the time 1-03-2007:10-06-2008}
\label{fig:fig3}
\end{figure}

\begin{figure}[h!]
\begin{center}
\includegraphics{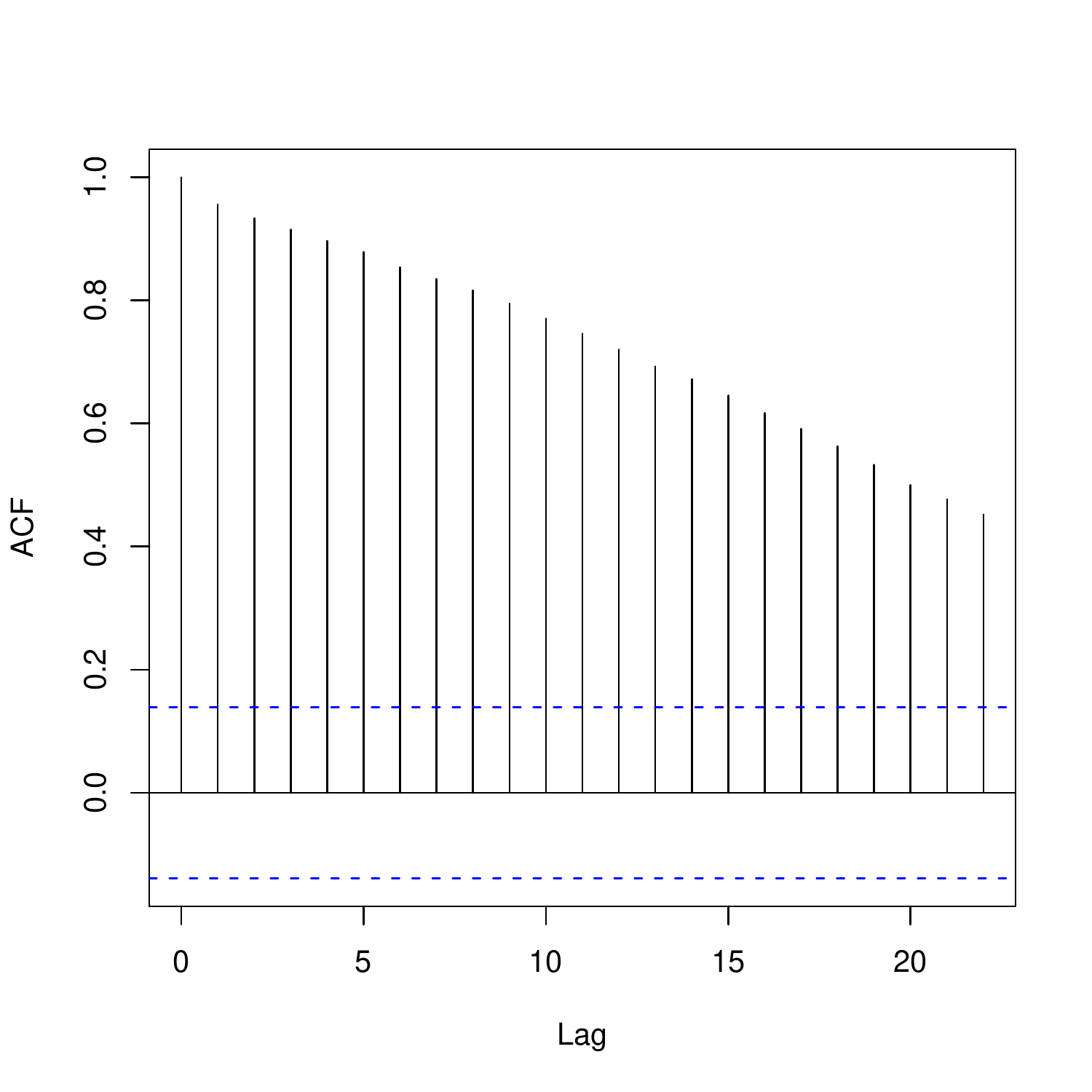}
\end{center}
\caption{Autocorrelation of regression residuals for ALSI on the time 11-06-2008:29-10-2009}
\label{fig:fig4}
\end{figure}

The study covers the period from 1 March 2007 to 10 November 2009. As seen in figure \ref{fig:fig1} (Appendix A) the USE ALSI trends upwards to over 1100 and then slumps to as slow as 600 points. \cite{mayanja07} test CAPM on the period 2006---2007 which is characterised by an upward trend. This is demonstrated in figure \ref{fig:fig2}~(Appendix A) where a regression line was inserted on the time series for the period 1st March, 2007 to 10th June, 2008---the date when USE ALSI attained its highest value of 1162.49 and then began falling. A regression line was also inserted for the period 11th June, 2008 to 29 October 2009. Sample sizes for the first and second regressions below are  183 and 198 respectively.
\begin{table}[ht]
\begin{center}
\caption{Linear Regression model for the period 1-Mar-2007:10-Jun-2008}
\label{tab:tab1}
\begin{tabular}{rrrrr}
  \hline
 & Estimate & Std. Error & t value & Pr($>$$|$t$|$) \\ 
  \hline
(Intercept) & 813.5209 & 8.3054 & 97.95 & 0.0000 \\ 
  t & 1.1144 & 0.0779 & 14.31 & 0.0000 \\ 
   \hline
\end{tabular}
\end{center}
\end{table}
\begin{table}[ht]
\begin{center}
\caption{Linear Regression model for the period 11-Jun-2008:10-Nov-2009}
\label{tab:tab2}
\begin{tabular}{rrrrr}
  \hline
 & Estimate & Std. Error & t value & Pr($>$$|$t$|$) \\ 
  \hline
(Intercept) & 1226.9543 & 36.6741 & 33.46 & 0.0000 \\ 
  t2 & -1.5107 & 0.1266 & -11.94 & 0.0000 \\ 
   \hline
\end{tabular}
\end{center}
\end{table}Figures \ref{fig:fig3} and \ref{fig:fig4} clearly show that the residuals are not white noise for both scenarios. Trend does not explain all the variation in the USE ALSI. However, the respective adjusted R-squareds are 0.52 and 0.42, which means that the effect of trend on the variation of the USE ALSI cannot be ignored. The time period was therefore chosen because not only is it a period marked with a lot of volatility, it also captures the trending of the USE, and most specifically the negative trend that has never been experienced on the USE.
\\
\\
The selected sample consists of 10 stocks of the 11 companies included in the USE ALSI. Ten (10) stocks only are considered, since the 11th was listed recently in June 2009. This consideration is adopted from \cite{bjs} who used all listed securities as of January 1932 for which atleast 24 months of previous monthly returns were available. 
\subsection{Data}

In this article, monthly stock returns from the 10 companies listed on the Uganda Stock Exchange for the period of March 2007 to November 2009, are used. The USE is just going through a computerisation phase, and as such there is no reliable database that can be accessed over the internet. The preparation of this article therefore required the writing of an application to mine the required data from the \texttt{pdfs} provided at \href{http://www.use.or.ug}{\texttt{http://www.use.or.ug}}. The application was written in \texttt{C++} using the \texttt{STL} library. The Standard Template Library (STL) is a library of C++ template classes for commonly occurring data structures (such as lists and vectors), algorithms (for example, sorting, searching and extracing information) as well as functionality for navigating in data structures. It is part of the ISO C++ standard and is not specific to a particular vendor. Thus, code that one writes and uses with a C++ compiler from vendor A will run using a C++ compiler from vendor B. Furthermore, the components in STL have been designed and implemented with performance in mind.\footnote{See \cite{duffy} for a detailed discussion}. The Daily price lists (pdfs) were converted into text files, and the application was run on these to \lq pick up\rq ~ the prices using the \verb$price_picker$ function. The source code and a sample text file are included in appendix B. 
\\
\\
All stock returns used in the study are adjusted for dividends as required by the CAPM. The USE ALSI, which is a market value weighted index comprising of the 11 stocks listed on the Uganda Stock Exchange, is used as a proxy for the market portfolio. 
\\
\\
The 91-day Ugandan Treasury Bill was used as the proxy for the risk-free asset. The treasury bill data was collected from Bank of Uganda research department and cross-checked with that available for download at~\href{http://www.bou.or.ug}{\texttt{http://www.bou.or.ug}}. In order to obtain the montly risk-free rates, the effective yield quoted for the treasury bill was de-annualized.
\section{Methodology}

Firstly, monthly stock returns were computed using 
 \begin{equation}
r_{it}= \log P_{it}-\log P_{i,t-1}
\label{eq:eq1}
\end{equation} 
where \[
\begin{array}{lcl}
 r_{it}& \mbox{---}&\mbox{logarithmic return of stock } i \mbox{ between time }$t-1$ and $t$\\
 P_{it}& \mbox{---}&\mbox{closing price of stock } i, \mbox{ on the 1st day of month } $t$,\\
 P_{i,t-1}& \mbox{---}&\mbox{closing price of stock } i, \mbox{ on the 1st day of month } $t-1.$ \\
 \end{array} \] The original CAPM by \cite{sharpe64} and \cite{lintner65} is given by 
 \\ 
 \begin{equation}
   \mu_v = r_f + (\mu_m-r_f)\beta_v
   \label{eq:eqn2}
   \end{equation}
   where \[
 \begin{array}{lcl}
 \mu_v &\mbox{---}&\mbox{expected return on stock }v,\\
 r_f &\mbox{---}&\mbox{risk-free rate of return},\\
 \mu_m&\mbox{---}&\mbox{expected return on market portfolio},\\
 \beta_v&\mbox{---}&\mbox{systematic risk of stock } v. \\
 \end{array} \] However, in this article, beta was estimated using historical returns and a proxy of the market portfolio, such that the relation being examined was not (\ref{eq:eqn2}) but 
 \begin{equation}
 (r_{it}-r_{ft}) = \gamma_0+(r_{mt}-r_{ft})\beta_i+ \epsilon_{it}
 \label{eq:eqn3}
 \end{equation}
as used by \cite{bjs}. Where, \[
 \begin{array}{lcl}
 r_{it}&\mbox{---}&\mbox{return on stock } i \mbox{ at time } t,\\
 r_{ft} &\mbox{---}&\mbox{rate of return on a risk-free asset,}\\
 \gamma_0&\mbox{---}&\mbox{zero-beta rate or alpha,}\\
 r_{mt} &\mbox{---}&\mbox{rate of return on the market index,}\\
 \beta_i&\mbox{---}&\mbox{estimate of beta for stock } i,\\
 \epsilon_{it}&\mbox{---}&\mbox{random disturbance term for stock } i \mbox{ observed at time } t.\\
 \end{array} \] Equation (\ref{eq:eqn3}) can be expressed using excess return notation: 
 \begin{eqnarray*}
  R_{it}& = &r_{it}-r_{ft} \\
  R_{mt}& = &r_{mt}-r_{ft}\\
  \end{eqnarray*}  Where $R_{it}$ is the excess return on stock $i$, and $R_{mt}$ is the excess return on the market index. Therefore, step two involved the computation of both excess returns, followed by the regression of $R_{it}$ on $R_{mt}$ for the 10 stocks.
\\
\\
The next step was to compute average portfolio excess returns. \cite{bjs} came up with a clever strategy that creates portfolios with very different betas for use in empirical tests. They estimate betas based on history, sort assets based on historical betas, group assets into portfolios with increasing betas, hold the portfolios for a selected number of years, and change the portfolio composition periodically. However, this strategy is not used in this article since most investors on the Uganda Stock Exchange hold portfolios with most if not all the stocks listed on the USE. Also, \cite{grigoris06} use 10 portfolios with 10 stocks each following the \cite{bjs} strategy, and as such one can argue that there are few stocks listed on the USE to proceed this way reliably. The 10 stocks are all put in one equally-weighted portfolio, and the average portfolio excess returns are given by 
\\
\\

   \begin{equation}
   R_t = \frac{\sum_{i=1}^{10} R_{it}}{10}
   \label{eq:eqn4}
   \end{equation}
  where, $R_{it}$ is the excess return of stock $i$ at time $t$. The analysis is devoid of selection bias since all the stocks are put in one portfolio. The following equation was used to estimate the portfolio beta. 
  
 \begin{equation}
 R_t = \psi_0+ R_{mt}\beta + \epsilon
 \label{eq:eqn5}
 \end{equation} where,
 \[ \begin{array}{lcl}
 R_t &\mbox{---}&\mbox{average excess return on the portfolio}, \\
 \psi_0 &\mbox{---}&\mbox{zero-beta rate}, \\
 R_{mt} &\mbox{---}&\mbox{market price of risk},\\
 \epsilon&\mbox{---}&\mbox{random disturbance}.
 \end{array} \]
The procedure above was repeated for monthly stock returns. The analysis proceeded by examining the R-squared values, and testing hypotheses about the zero-beta rate and beta coefficient. 
\\
\\
Emphasis was put on both hypotheses tests\footnote{See \cite{bjs}and \cite{famafrench}. \cite{banz81} used the absolute value of the t-statistic to conclude that the size effect is large and statistically significant.} and R-squared values, as has always been in CAPM analysis. The hypotheses tests carried out were:
\begin{itemize}
\item $\psi_0\neq 0$ or zero beta rate is not equal to zero.
\item $\beta>0$ or there is a positive price of risk in the capital market.
\end{itemize}
All the computation was carried out in \texttt{R}.
\section{Results and Analysis}
\begin{figure}[h!]
\begin{center}
\includegraphics{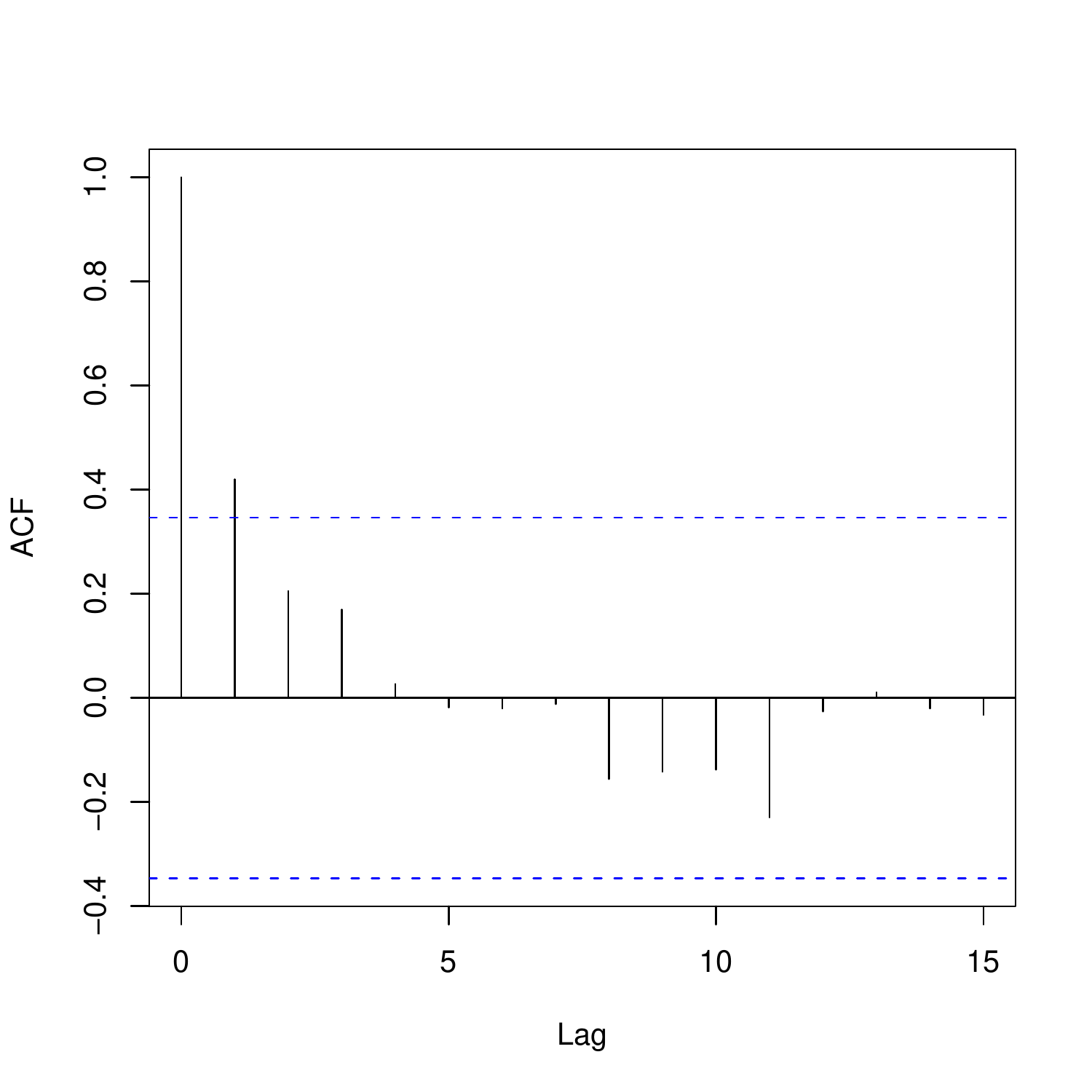}
\end{center}

\caption{Autocorrelation of the residuals obtained from the regression of excess portfolio return on the excess return on the market index}
\label{fig:fig5} 
\end{figure}

Using monthly stock returns, the excess return of each stock was regressed on the excess return on the market index. Table \ref{tab:tab3} shows the estimated stock beta coefficients. Sample size for regression = 32.
\\
\begin{table}[h!]
\begin{center}
\caption{Stock beta estimates obtained using monthly stock returns}
\label{tab:tab3}
\begin{tabular}{lcrcc} \hline
Stock Name  & Estimated Beta  & t-value & Std. Error  & R-squared\\ \hline
BATU    & 0.1168   & 0.324 &0.36&0.0035  \\ 
BOBU    & 0.6305  & 0.974  &0.647&0.0307 \\ 
DFCU    &0.1796  &0.781    &0.23&0.0199\\
EABL    &1.2645  &15.468 &0.082 &0.8886\\
JHL    &0.9597   &3.803  &0.252 &0.3252\\
KA      &1.2629    &7.604  &0.166 &0.6584\\
KCB   &1.2782   &7.159 &0.179  &0.8506\\
NVL   &0.4137  &1.252  &0.33  &0.0496\\
SBU   &0.3629   &1.869  &0.194 &0.1043\\
UCL  &0.9216  &0.55  &1.675   &0.01\\ \hline
\end{tabular}
\end{center}
\end{table}
\\
The estimated beta coefficients range from 0.1168 to 1.2782. Five out of the ten stocks\rq ~beta coeffecients are positive and statistically significant at the 5\% level. CAPM predicts that higher risk is associated with higher return, and therefore we expect positive beta estimates for CAPM to hold, and yet five stocks have beta coeffecients that are not statistically different from zero, even at the 10\% level. 
\\
\\
The R-squared values, as seen in table \ref{tab:tab3}, are also generally very low. It is only the EABL, KA and KCB stocks that have their variation in excess return fairly explained by the excess return on the market index. This is equivalent to having betas that are fairly efficient in explaining the relation between market risk and return, for only three of the five stocks with positive statistically significant beta.  The R-squared value is the ratio of market risk to the sum of market and firm-specific risk, and as such a low value points to the inefficiency of beta---the measure of market risk. 
\\
\\
The article tests the \cite{bjs} CAPM version, which requires the existence of a zero beta rate in the CAPM equation(\ref{eq:eqn2}), and yet as seen from table \ref{tab:tab4}, it is only the KA stock with a zero beta rate that is statistically different from zero at the 10\% signifance level. 
\\
\begin{table}[ht]
\begin{center}
\caption{Stock zero beta rate estimates using monthly stock returns}
\label{tab:tab4}
\begin{tabular}{lcrcc} \hline
Stock Name  & Estimated zero beta    & t-value          & Std. Error              & p-value\\ \hline
BATU    & -0.0128   & -0.414 &0.031&0.682  \\ 
BOBU    & -0.0354   & -0.634 &0.056&0.531 \\ 
DFCU    &6e-04    &0.028  &0.02&0.978\\
EABL    &0.0102    &1.453  &0.007&0.157\\
JHL     &-0.0247     &-1.137   &0.022 &0.2647\\
KA      &-0.034      &-2.375    &0.014  &0.0241\\
KCB     &0.0072     &0.368   &0.02 &0.722\\
NVL     &-0.0018     &-0.064   &0.028 &0.95\\
SBU     &0.0015     &0.092   &0.017 &0.9276\\
UCL     &-0.1088     &-0.753   &0.144 &0.457\\ \hline
\end{tabular}
\end{center}
\end{table} 
\\
Two serious issues have been brought out by the analysis --- failure to have statistically significant beta coeffecient and zero beta rate for most of the stocks on the USE. Does this therefore mean that CAPM does not hold on the USE? Not necessarily. Individual stocks are affected by random noise much more than portfolios, as discussed by \cite{jaganna}, which is why \cite{bjs} came up with an ingenious strategy that creates portfolios with very different betas for use in empirical tests. 
\\
\\
However, the \cite{bjs} portfolio strategy was not used in this article, as has been argued. The average excess return of the portfolio created by pooling up all the stocks, was regressed on the excess return on the market index, and the results obtained are shown in table \ref{tab:tab5}. Sample size for regression = 32.
\\
\begin{table}[ht]
\begin{center}
\caption{Portfolio beta estimate}
\label{tab:tab5}
\begin{tabular}{cccc} \hline
Estimated Beta & t-value & Std. Error & R-squared  \\ \hline
0.6832 & 3.512 &0.1946&0.2913 \\ \hline
\end{tabular}
\end{center}
\end{table}
\\
The portfolio beta coefficient is statistically significant at the 1\% level, which strongly supports the CAPM prediction that higher risk is associated with higher return. Figure \ref{fig:fig5} (Appendix A) shows that the serial correlation of the residuals obtained from the regression of the excess portfolio return on the excess market return is 0.4, with the Durbin-Watson statistic being approximately 1.2. As a rule of thumb, if Durbin-Watson is less than 1.0, there may be cause for alarm about the bias in the OLS standard errors and test statistics, which is not the case here. However, the R-squared value is low, which points to the ineffectiveness of the beta coefficient in explaining the relationship between risk and return. 
\\
\\
The other hypothesis tested was whether the zero-beta rate is not equal to zero or not. Table \ref{tab:tab6} below has the $t$ value used in the test.
\\
\begin{table}[ht]
\begin{center}
\caption{Estimated zero beta rate}
\label{tab:tab6}
\begin{tabular}{ccc} \hline
Est.zero beta rate       & t-value & Std. Error \\ \hline
-0.0203  & -1.213 & 0.0168 \\ \hline
\end{tabular} 
\end{center}
\end{table}
\\
The zero-beta rate is not statistically significant at the 10\% level. Using the data from 1st March 2007 to 29th October 2009, the study cannot reject the hypothesis that the zero-beta rate $\gamma_0$ is equal to zero. 
\\
\\
The study set out to find out whether the CAPM holds on the USE, and specifically tested \cite{bjs} CAPM version on the USE. The results obtained show that it is only the beta coefficient that is statistically significant.

\section{Conclusion}
There isn't sufficient evidence for the \cite{bjs} CAPM version, since the zero-beta rate is not statistically different from zero at the 10\% level. Therefore, the traditional form of CAPM represented by equation(\ref{eq:eqn2}) holds on the USE. However, the R-squared value of 0.2913 shows that the variation in excess portfolio returns is not well explained by the excess return on the market index. This means that the beta coeffecient does not offer a good explanation about the relationship between return and systematic risk on the USE.
\\
\\
The estimated portfolio beta coefficient has a value of 0.6832 which is less than 1, showing that the systematic risk on the USE is low. It is likely that portfolios on the USE do not offer higher risk-adjusted returns. This finding is consistent with the fact that most emerging markets are characterised by low risk.
\\
\\
A growing number of studies have found that the cross-sectional variation in average security returns cannot be explained by the market beta alone. The most important among these are \cite{banz81}, \cite{rrl85} \& \cite{chan91}, and \cite{basu83} who respectively argued that size, book-to-market value ratio, macroeconomic variariables and the price-to-earnings ratio account for a sizeable portion of the cross-sectional variation in expected return.  Further work would therefore involve the testing the effect of such variables following the analysis of for example \cite{famafrench}.

\section*{Acknowledgements}
I am most grateful to Dr. Colin Rowat (University of Birmingham), Mr. Daniel Nvule and Dr. Henry Mwambi (Kwazulu-Natal) for their comments and insightful advice. 
\pagebreak
\ifx\undefined\BySame
\newcommand{\BySame}{\leavevmode\rule[.5ex]{3em}{.5pt}\ }
\fi
\ifx\undefined\textsc
\newcommand{\textsc}[1]{{\sc #1}}
\newcommand{\emph}[1]{{\em #1\/}}
\let\tmpsmall\small
\renewcommand{\small}{\tmpsmall\sc}
\fi

\pagebreak

\noindent\textbf{\Large Appendix }
\begin{appendix}

\section{Figures}

\noindent (\texttt{R-sweave}, \LaTeX)
\setkeys{Gin}{width=5in, height=2.5in}

\setkeys{Gin}{width=5in, height=2in}

\section{Data Minning Application: \texttt{use\_analysis.exe}}

\subsection{Header file---\texttt{use\_analysis.hpp}}
\begin{verbatim}
#ifndef US_ANALYSIS_HPP
#define US_ANALYSIS_HPP
//contains the price_picker function which picks 
//the date, current ALSI and
//current and previous prices from the text files.

//David Wakyiku 2009


#include <iterator>
#include <vector>
#include <fstream>
#include <iostream>
#include <strstream>
#include <string>
#include <algorithm>

using namespace std;

//listed company names in string array
 string comps[]={"DATE:","CURRENT", "BATU", "BOBU","DFCU","EABL", "JHL",
 "KA","KCB","NVL", "SBU","UCL"}; //line markers in the txt file from 
               //which data is to be mined.
 string months[]={"Jan","Feb","Mar","Apr","May", "Jun","Jul","Aug",
 "Sep", "Oct","Nov","Dec"};
 string days[]={"01","02","03","04","05","06", "07","08","09","10",
 "11", "12","13","14",  "15","16","17","18","19","20","21","22","23",
 "24","25", "26","27","28","29","30", "31"};
  string years[]={"2007","2008","2009"}; 
  //period of interest is March 2007
  //to October 2009
 vector<string> compns(comps,comps+12);
 vector<string>vmonths(months,months+12);
  vector<string> vyears(years,years+3);
 vector<string> vdays(days,days+31);
const char* delimiters ="+,k*>\"#^_<;'";

//PRICE_PICKER function to pick the prices from the text file 
 vector<string> price_picker(const char* filename)
 {
int count=0; //read the first 23 lines of the file
int k =0;
string s,line;
vector<string>::iterator vit;
vector<string> line_holder;
//line_holder holds line data as it comes form the file
vector<string> ldata;
 //ldata holds the wanted data..after filtering operation 
 //have been carried out on line_holder below.
ifstream in(filename);
if(in)
while(getline(in,line)&&count<23)
{ //data contained in first 23 lines
  istrstream strsplit(line.c_str()); 
     //split the line into "words"
   while(strsplit>>s)
   {
    while(k!=string::npos)
     { //remove the commas and * ==delimiters
       k = s.find_first_of(delimiters,k+1);   
       if(k!=string::npos) s.erase(k,1);
      }          
      line_holder.push_back(s);
       k=0; 
        //for each portion, we start the search at the beginning.
     }  
              
      vit= find_first_of(line_holder.begin(),line_holder.end(),
        compns.begin(),compns.end()); 
       //find first_of line in  compns vector
       //this is to determine the line we want to push into ldata
       //compns has the line markers for the lines of interest
         if(vit!=line_holder.end())
         { //begin if1
                                               
          if(*vit==line_holder[0])
           { //begin if2
              if(*vit==compns[0])  //Pick date 
               ldata.push_back(line_holder[1]);
 
               else
               {
               if(*vit==compns[1]) //Picking current ALSI
                 {
                     if(line_holder.size()<3)
                      ldata.push_back(string("000"));
                                    
                       else
                        ldata.push_back(line_holder[2]);
                   }
                               
                 if(*vit!=compns[1])
                   {
                       ldata.push_back(line_holder[4]); 
                          //pick the close and open price
                          // ldata.push_back(line_holder[5]);
                     }
                  } //end of else
           }//end of if2
           }//end of if 1
           line_holder.clear(); //line_holder has to clear
            count++;
                  
}//end of mother-while loop
                
in.close();
return ldata;
}    //end of PRICE_PICKER function    
   
#endif 
\end{verbatim}

\vspace*{.3in}

\subsection{Source file---\texttt{use\_analysis.cpp}}

\begin{verbatim}
#include "us_analysis.hpp"

//David Wakyiku 2009

int main()
{
    
string txt_file_name;
vector<string>::iterator it1,it2,it3;   
vector<string> data; //relevant data
ofstream out("closing_px.out"); //closing_px.out stores the "picked" prices.


/* The text files are named the same way. the name is of the format
USE_MARKET_REPORTmonth_date-year. The for loop goes through the folder with 
the text files and picks all the prices using the price_picker function
defined in us_analysis.hpp

*/
for(it3=vyears.begin();it3!=vyears.end();it3++)
{
for(it1=vmonths.begin();it1!=vmonths.end();it1++)
{
   for(it2=vdays.begin();it2!=vdays.end();it2++)
   {
     txt_file_name= "USE_MARKET_REPORT"+*it1+"_"+*it2+"-"+*it3+".txt";
                 
     const char* name = txt_file_name.c_str();
             
           
     data =  price_picker(name);    //picks the prices in the file       
     if(data.size()!=0)
     {
      copy(data.begin(),data.end(),ostream_iterator<string>(out,"\t"));
      out<<endl;
      }
    }//inner for loop
  } //outer for loop
}//outest for loop
out.close();
system("pause");
return(0);
}          
     
     
    
    
    
    
    
\end{verbatim}

\newpage

\subsection{Sample text file}

\begin{figure}[h!]
\begin{center}
\includegraphics[width=12cm, height =12cm]{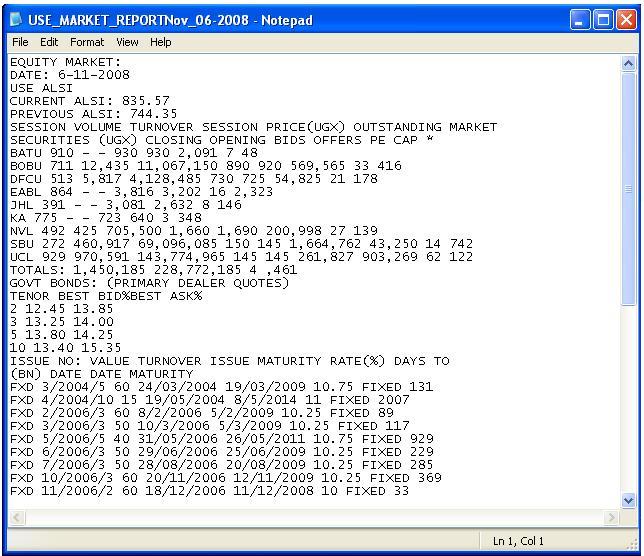}

\caption{Sample of the text files onto which the data mining application --- \texttt{use\_analysis.exe} is run} 
\label{fig:fig6}
\end{center}
\end{figure}

\newpage

%

\end{appendix}
\end{document}